\documentclass[]{elsart}

\usepackage[dvips]{graphics, color}   
\usepackage{latexsym}
\usepackage{amsmath}
\usepackage{amssymb}
\usepackage{rotating}

\begin{document} 
\begin{frontmatter}

\title {Early Results on Radioactive Background Characterization for Sanford Laboratory and DUSEL Experiments }
\thanks[*]{This work is funded by NSF PHY-0758120}
\author[a]{D.-M. Mei},
\ead{Dongming.Mei@usd.edu}
\author[a,b]{C. Zhang},
\author[a] {K. Thomas},
\author[c] {F. Gray}
 
\address[a]{Department of Physics, The University of South Dakota, Vermillion, SD 57069}
\address[b]{College of Sciences, China Three Gorges University, Yichang 443002, China}
\address[c] {Department of Physics and Computational Science, Regis University, Denver, CO 80221}
\begin{abstract}
Measuring external sources of background for a deep underground laboratory at the Homestake Mine is an important step for the planned low-background experiments. The naturally occurring $\gamma$-ray fluxes at different levels in the Homestake Mine are studied using NaI detectors and Monte Carlo simulations. A simple algorithm is developed to convert the measured $\gamma$-ray rates into $\gamma$-ray fluxes. A good agreement between the measured and simulated $\gamma$-ray fluxes is achieved with the knowledge of the chemical composition and radioactivity levels in the rock. The neutron fluxes and $\gamma$-ray fluxes are predicted by Monte Carlo simulations for different levels including inaccessible levels that are under construction for the planned low background experiments.   
\end{abstract}

\end{frontmatter}

\section{Introduction}

Neutrinos and dark matter are believed to hold the key to physics beyond the Standard Model~\cite{sep,sej,fky, asb, gju, mwg, bwl}. However, very little is known about the general properties of neutrinos such as their absolute mass and magnetic moment, whether they are Dirac or Majorana particles, or the number of species. Even less is known about the nature and quantity of dark matter in the universe.  A deeper understanding of both neutrinos and dark matter is important to understand physics beyond the Standard Model. Planned underground low-background experiments are intended to probe these properties using rare event physics processes such as neutrinoless double-beta decay, neutrino oscillations, and the direct detection of dark matter. Because of a very low event rate, the reduction of backgrounds from radioactivity and cosmic rays is essential to these experiments. 

The solution to reducing cosmic rays is to build experiments in a deep underground laboratory. However, deep underground experiments require appropriate shielding against $\gamma$ rays and neutrons from natural radioactivities in the surrounding materials and rocks. As the major contributors to the radiation, the $\gamma$ rays and neutrons range from keVs to MeVs in kinetic energy. The $\gamma$ rays are mainly from the decays of $^{238}$U, $^{232}$Th, and $^{40}$K in the rocks. The neutrons come primarily from ($\alpha$,n) reactions induced by $^{238}$U, and $^{232}$Th decays and fission decays of $^{238}$U~\cite{meichao, heaton}. Care must be taken to understand the $\gamma$-ray and neutron fluxes as a function of energy in order to design appropriate shielding. This is usually done by measurements in combination with Monte Carlo simulations. SNO~\cite{sno}, Gran Sasso~\cite{gran, bell},Frejus~\cite{frej}, Boulby~\cite{boulby}, and other underground laboratories~\cite{baksan} have reported respective $\gamma$-ray and neutron fluxes in terms of the measurements and Monte Carlo simulations.
\par
The Homestake Mine in Lead, South Dakota, has been selected by the National Science Foundation (NSF) as the potential site for Deep Underground Science and Engineering Laboratory (DUSEL). As funding for DUSEL is being secured, an early science program is being developed at the interim Sanford Laboratory. This makes the characterization of the $\gamma$ rays and neutrons in the Homestake Mine a necessary endeavor. Generally, the fluxes of $\gamma$ ray and neutrons reveal the concentrations of the radioactive sources and the composition of the surrounding rocks. Because the geology of Homestake Mine varies from location to location, the variations in the radioactivity concentration in the rock~\cite{geology} result in site-dependent $\gamma$-ray and neutron fluxes.  
\par
We demonstrate for the first time the site-dependent $\gamma$-ray flux with measurements at different levels using NaI detectors in Section 2. We measure the $\gamma$-ray fluxes at the surface in the administration building, the underground 800-ft level, 2000-ft level, and 4550-ft level. Several measurements at different spots separated by no more than 100 feet were done at the same level. Since many rock samples in the different levels at Homestake Mine 
have been collected and analyzed~\cite{AlSmith}, we are able to perform a Monte Carlo simulation to estimate the $\gamma$-ray fluxes for different levels. The simulated results are discussed and compared to the measurements in Section 3.

The neutron fluxes at different levels are characterized through a Monte Carlo simulation. The neutron yield and energy spectrum in the rock are calculated utilizing the tools and methods described in Ref.~\cite{meichao}. Those neutrons are tracked through the rock to the experimental hall. We elaborate on the neutron fluxes in Section 4.

\section{The $\gamma$-ray fluxes at different levels}

The $\gamma$ rays from natural radioactivity in the rocks at the Homestake Mine are measured using NaI detectors. Three identical NaI detectors were deployed to directly
measure $\gamma$ rays at different levels. The NaI detectors are $3^{''}\times 3^{''}$ from Bicron Crystal Corporation. All three detectors are read out using a Digital Gamma Finder Four Channel (DGF4C) CAMAC module supplied by X-Ray Instrumentation Associates (XIA)~\cite{xia}. This module is a 14-bit digitizer with 40 MHz sampling rate. The CAMAC create is connected to an Express card of a Hewlett Packard (model 8730) laptop running Microsoft Windows Vista. The system was controlled using the standard software provided by XIA. This data acquisition software runs in the IGOR Pro environment~\cite{igor} and produces binary data files that are read in and analyzed using the ROOT framework~\cite{root}.

Three detectors were arranged horizontally in the same plane and were separated by about 50 cm from each other. The measured $\gamma$-rays energy spectrum from each detector provides a cross-check to spot anomalous detector response, which can then be corrected on the site. The internal background from the detectors was measured at the surface lab with lead bricks that surround the detectors. The contribution from the internal background to the energy range of interest (less than 3 MeV) 
is less than 0.1\% of the measured external $\gamma$ rays.
\par
The detectors were calibrated using a $^{60}$Co $\gamma$-ray source. This 
simple calibration allows us to understand the energy response and identify the well-known peaks from the external radiation. The four well-known peaks, 0.609 MeV and 1.764 MeV from $^{214}$Bi in $^{238}$U decay chain, 1.461 MeV from $^{40}$K decay, and 2.615 MeV from $^{232}$Th decay chain, are distinguished. These four peaks are used as the calibration points for each collected data file. Despite the small variations in the gain of the detectors resulting in an energy shift in the spectrum, the four well-known peaks always provide a reliable self-calibration to the entire measured spectrum. 

The differential flux is determined by using  
the formula 
\begin{equation}
\Phi (E) =\frac{R} {\sigma(E) \cdot m},
\end {equation}
where $\Phi (E)$ is  in unit of cm$^{-2}$s$^{-1}$keV$^{-1}$, $R$ is the registered count rate (counts/s) of the detector for a selected energy region, $\sigma$(E) is the photon attenuation factor (cm$^{2}$/g) of NaI with respect to the selected energy~\cite{attu}, and $m $ is the effective mass (g) of the NaI crystal. The term $\sigma$(E) (cm$^{2}$/g) $\cdot$ $m$(g) accounts for the effective area of the detector that subtends the $\gamma$ rays and the detection efficiency of the $\gamma$ rays as a function of energy. The value of this term is obtained by the combination of the calibration and Monte Carlo simulation.

\par
\begin{figure}
\includegraphics[width = \textwidth]{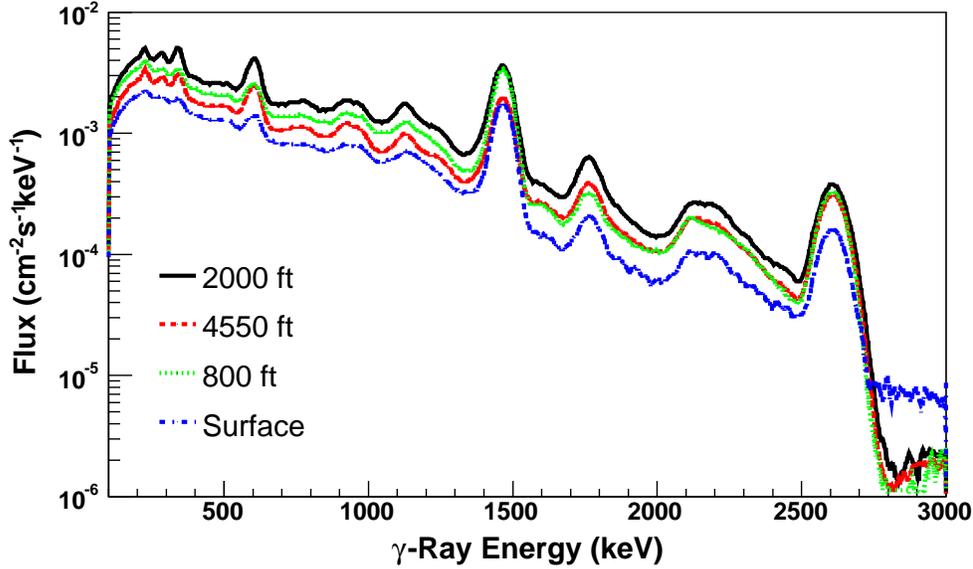}
\caption{\label{Fig_flux}
Shown is the measured differential $\gamma$-ray flux at the surface administration building and underground on the 800-ft, 2000-ft, and 4550-ft levels from one of the three detectors. }
\end{figure}

The final experimental results for the surface administration building and
underground on the 800-ft, 2000-ft and 4550-ft levels are shown in Figure\ref{Fig_flux}. The $\gamma$-ray flux with energy greater than 0.1 MeV are listed in Table \ref{Tab_flux}. These numbers are the average of the three measurements. The variation in the $\gamma$-ray flux between the three detectors with a distance of 50 cm from each other is about 3\%. 
\begin{table}
\caption{The integrated $\gamma$-ray flux at the surface administration building,
underground on the 800-ft, 2000-ft and 4550-ft levels. The quoted errors are the largest differences between the average of the three detectors and the individuals.   }
\centering
\begin{tabular}{ccccc}\hline
 & \multicolumn{4}{c}{The measured $\gamma$-ray flux (cm$^{-2}$s$^{-1}$)} \\
 & E$>$0.1 MeV & E$>$1 MeV & E$>$2 MeV & E$>$3 MeV \\ 
 Surface& 1.56$\pm$0.05& (4.63$\pm$0.14)$\times$10$^{-1}$& (5.52$\pm$0.17)$\times$10$^{-2}$& (1.09$\pm$0.03)$\times$10$^{-3}$ \\ 
 800 ft& 2.65$\pm$0.08& (7.97$\pm$0.24)$\times$10$^{-1}$& (9.49$\pm$0.28)$\times$10$^{-2}$& (4.81$\pm$0.20)$\times$10$^{-4}$ \\ 
 2000 ft& 3.42$\pm$0.10& (1.04$\pm$0.03)& (1.26$\pm$0.04)$\times$10$^{-1}$& (7.05$\pm$0.21)$\times$10$^{-4}$\\ 
 4550 ft& 2.16$\pm$0.06& (6.32$\pm$0.19)$\times$10$^{-1}$& (9.64$\pm$0.29)$\times$10$^{-2}$& (6.01$\pm$0.18)$\times$10$^{-4}$ \\ 
\hline
\end{tabular}
\label{Tab_flux}
\end{table}
It is worthwhile mentioning that the contribution to the $\gamma$-ray flux from radon is estimated to be less than  1\% for a radon level as high as 500 Bq/m$^{3}$, which is an example of the radon level measured at Homestake~\cite{meikeenan}. Therefore, the difference in the radon levels can not explain the difference seen in the measured $\gamma$-ray fluxes. 
The difference in the $\gamma$-ray fluxes measured for different levels corresponds to  the variation of the radioactivity levels in the rocks near the experimental sites. The variation in the radioactivity levels is associated to the geologic formations for different levels as demonstrated via the Monte Carlo simulation described in Section 3.

\section{Simulated $\gamma$-ray flux for different levels at the Homestake Mine}

\begin{figure}
\includegraphics[width = \textwidth]{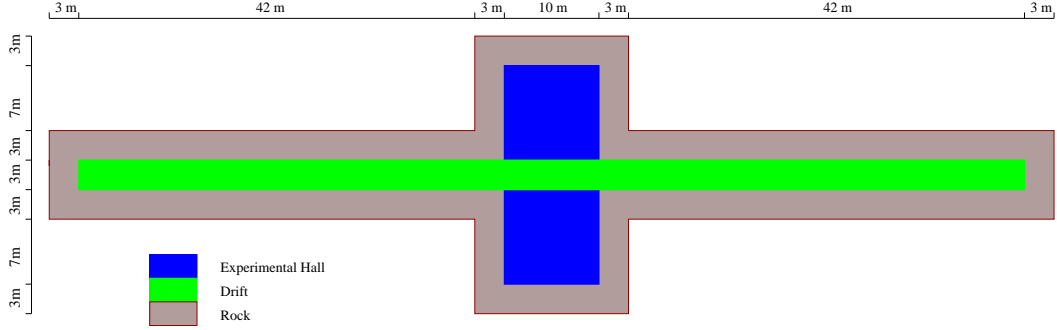}
\caption{\label{geometrymap} 
Sketch of the geometry in the simulation showing two experimental halls intersected by a drift surrounded by 3 m of rock. }
\end{figure}

The measured $\gamma$ rays are primarily from the decay 
of actinide elements in the surrounding rocks. The contamination by 
the actinide elements and the rock composition must be taken into account in the Monte Carlo simulation. Since the highest $\gamma$-ray energy from natural radioactivity is 2.615 MeV and the attenuation length of such a $\gamma$ ray in the Homestake rock is about 9.4 cm, a rock thickness of 300 cm is sufficient to simulate the $\gamma$ rays from the rock. On the other hand, the attenuation length of a $\gamma$ ray with 2.615 MeV energy in air is about 21 meters. Therefore, we define the simulation geometry as below (Figure~\ref{geometrymap}): 1) the sensitive detection areas are two experimental halls 10 m$\times$ 10 m$\times$ 3 m in dimension; 2) these two halls are separated by a 100 m$\times$ 3 m$\times$ 3 m drift in the middle; 3) both the hall and the drift are filled with standard air; and 4) the outer space of the experimental hall and the drift are surrounded by a 3 m thickness of rock. 

A Homestake rock sample 278-2 \cite{rocksample} is applied in our simulation with the chemical composition given in Table \ref{composition}. The chemical composition varies from location to location at different levels. Twenty rock samples near the surface were collected and analyzed~\cite{rocksample}. The rock chemical composition affects the density and therefore the attenuation of the $\gamma$ rays through the rock. The largest variation in chemical composition seen in the samples was used to estimate the variation in the $\gamma$-ray fluxes. This variation results in a maximum uncertainty of about 10\% in the simulated $\gamma$-ray fluxes. It is clear that the chemical composition cannot account for the variation seen in the $\gamma$-ray fluxes at different levels. 
\begin{table}
\caption{The chemical composition of the Homestake rock sample used in the simulation~\cite{rocksample}.}
\centering
\begin{tabular}{|l|c|}\hline
Rock Component &Composition  \\
(Sample 278-2)& (\% weight)  \\
\hline
SiO$_{2}$& 43.7$\pm$0.4\\
TiO$_{2}$& 1.22$\pm$0.01\\
Al$_{2}$O$_{3}$& 13.6$\pm$0.1\\
FeO& 12.7$\pm$0.1\\
MnO& 0.13$\pm$0.01\\
MgO& 7.0$\pm$0.1\\
CaO& 7.9$\pm$0.1\\
Na$_{2}$O& 2.87$\pm$0.03\\
K$_{2}$O& 0.21$\pm$0.002\\
P$_{2}$O$_{5}$& 0.07$\pm$0.001\\
H$_{2}$O& 10.7$\pm$0.01\\
\hline
\end{tabular}
\label{composition}
\end{table}
\par
The major contributions to the measured $\gamma$ rays are from $^{238}$U, $^{232}$Th, $^{40}$K and their daughters. The $\gamma$-ray decay ratios and energies from the decay chain of these isotopes were studied in detail in Ref.~\cite{branch}. The simulation is performed by using 
the GEANT4 simulation package~\cite{geant4}. The $\gamma$ rays in a few MeV range can be simulated in an accuracy of less than 1\%~\cite{geant4}. The dominated uncertainty of about 10\% comes from the variation of the chemical composition in the rock. The input $\gamma$ rays are uniformly generated in the rock with the emitted energies from the decay chains.
We summarize the simulated $\gamma$-ray fluxes in Table \ref{1ppm-gamma-flux}. 
\begin{sidewaystable}
\caption{The $\gamma$-ray flux in the simulated experimental hall from natural radioactivity in the rock. 
The rows indicate the energy of $\gamma$ ray in MeV size bins. The columns represent the further subdivision of the data 
in the MeV bins into 0.1 MeV bins. The uncertainty in the simulation is about 10\%.}
\centering
\begin{tabular}{l l|ccccccccccc}\hline
\multicolumn{2}{l}{Range (MeV)} &$0- .1$ &$.1- .2$
                                &$.2- .3$ &$.3- .4$ &$.4- .5$ &$.5- .6$
                                &$.6- .7$ &$.7- .8$ &$.8- .9$ &$.9- 1$ &Sum \\
\hline
 $E_{\gamma}$ & Source & \multicolumn{10}{c}{
                           $\gamma$-ray flux (ppm$^{-1}$cm$^{-2}$s$^{-1})$}  \\
\hline
&$^{238}$U
&9.1e-2&1.4e-1&6.5e-2&4.1e-2&1.6e-2&1.1e-2&2.5e-2&7.6e-3&5.0e-3&5.2e-3&4.0e-1\\
$0$&$^{232}$Th
&3.5e-2&5.5e-2&3.3e-2&1.2e-2&6.7e-3&1.2e-2&3.0e-3&5.3e-3&2.7e-3&7.5e-3&1.7e-1\\
&$^{40}$K
&6.6e-2&1.4e-1&7.1e-2&3.4e-2&2.3e-2&1.9e-2&1.5e-2&1.2e-2&1.2e-2&1.1e-2&4.0e-1\\
\hline
&$^{238}$U
&5.7e-3&1.1e-2&7.4e-3&6.1e-3&4.1e-3&1.7e-3&1.0e-2&1.4e-2&1.5e-3&2.2e-4&6.2e-2\\
$1$&$^{232}$Th
&5.0e-3&1.4e-3&5.8e-4&5.1e-4&4.3e-4&1.4e-3&8.0e-4&2.2e-4&2.9e-4&1.4e-4&1.1e-2\\
&$^{40}$K
&1.2e-2&1.3e-2&9.4e-3&8.7e-3&1.6e-1&0.0e+00&0.0e+00&0.0e+00&0.0e+00&0.0e+00&2.0e-1\\
\hline
&$^{238}$U
&0.0e+00&1.5e-3&3.5e-3&0.0e+00&4.4e-4&0.0e+00&0.0e+00&0.0e+00&0.0e+00&0.0e+00&5.5e-3\\
$>2$&$^{232}$Th
&2.9e-4&6.5e-4&2.2e-4&4.3e-4&1.4e-4&7.2e-4&9.3e-3&0.0e+00&0.0e+00&0.0e+00&1.2e-2\\
&$^{40}K$
&0.0e+00&0.0e+00&0.0e+00&0.0e+00&0.0e+00&0.0e+00&0.0e+00&0.0e+00&0.0e+00&0.0e+00&0.0e+00\\
\hline
\hline
\end{tabular}
\label{1ppm-gamma-flux}
\end{sidewaystable}
\par
The simulated $\gamma$-ray flux is quoted in the number of $\gamma$ rays per ppm per cm$^{2}$ per second. The ppm stands for parts per million describing the concentration of radioisotopes in the rock, which allows us to calculate the $\gamma$-ray flux 
with the known concentrations of $^{238}$U, $^{232}$Th and $^{40}$K. 
There are three main rock formations at the Homestake Mine: Poorman Formation, Homestake Formation, and  Ellison Formation~\cite{geology, types}. Table \ref{rocktype} shows these three types of rock.
Taking the average values of the concentrations,  we obtain the 
$\gamma$-ray flux $\Phi$ (E $>$ 0.1 MeV) = 3.9$\pm$0.4, 2.3$\pm$0.2, and 9.5$\pm$1.0 cm$^{-2}$s$^{-1}$ for the Poorman, Homestake, and Ellision Formations, respectively. In addition, rhyolite intrusive rocks occur as thin dikes in the Homestake rock~\cite{geology} at different levels. The $^{238}$U, $^{232}$Th, and $^{40}$K contents in the rhyolite dikes are higher than these in the above three formations. The counted rhyolite samples show a maximum of 9.4 ppm of $^{238}$U, 12.2 ppm of $^{232}$Th, and 3.98 \% of K~\cite{geology}. However, the rhyolite dikes do not make up large volumes in the underground~\cite{geology}.
\begin{table}
\caption{The three main rock types and the contamination of 
 $^{238}$U, $^{232}$Th and $^{40}$K in the rocks~\cite{types} as measured from a number of samples.}
\centering
\begin{tabular}{|l|l|c|c|c|}\hline
& & $^{238}$U & $^{232}$Th & nat-K  \\
\hline
&Average &3.43 ppm &7.11 ppm  &2.66  \%\\
&Standard Deviation &3.73 ppm &5.64 ppm &1.05 \% \\
Poorman Formation&Number of samples & 7& 7& 141 \\
&Maximum Value &11.7 ppm &18 ppm &4.98 \% \\
&Minimum Value &0.1 ppm &0  ppm&0.04  \%\\
\hline
&Average &1.51 ppm &7.38 ppm &0.96 \% \\
&Standard Deviation &0.94 ppm &4.04 ppm &0.91  \%\\
 Homestake Formation&Number of samples &24 &24 &263 \\
&Maximum Value &3.5 ppm &15 ppm &4.31  \%\\
&Minimum Value &0.4 ppm &1.9 ppm &8.3e-3 \% \\
\hline
&Average &7.45 ppm &30.5 ppm &3.32  \%\\
&Standard Deviation &3.65 ppm  &16.5 ppm &0.95 \% \\
 Ellison Formation&Number of samples &2 &2 &55 \\
&Maximum Value &11.1 ppm &47 ppm &6.22  \%\\
&Minimum Value &3.8 ppm &14 ppm &1.16  \%\\
\hline
\end{tabular}
\label{rocktype}
\end{table}

To compare with the measured $\gamma$-ray fluxes at the 800-ft, 2000-ft, and 4550-ft levels, one must know the combination of the rock formation for these levels. A detailed geological study is under way. Using the preliminary knowledge from Refs.~\cite{ geology, tom}, we assume that the rock near the measurement locations at the 800-ft level is mainly dominated by the Poorman (80\%) and Homestake (20\%) Formations. This combination predicts a $\gamma$-ray flux (E$>$0.1 MeV) of 2.6$\pm$0.78 cm$^{-2}$s$^{-1}$, which is in good agreement with the measured 2.65$\pm$0.80 cm$^{-2}$s$^{-1}$ at this level. The rock close to the measurements at the 2000-ft level constitute mainly Homestake (80\%) and Ellison (20\%) Formations. The predicted $\gamma$-ray flux (E$>$0.1 MeV) for the 2000-ft level is about 3.7$\pm$1.1 cm$^{-2}$s$^{-1}$. Again, it is in a good agreement with the measured flux of 3.42$\pm$1.0 cm$^{-2}$s$^{-1}$. The rock adjacent to the measurements at the 4550-ft level is in the Homestake Formation. The predicted  $\gamma$-ray flux (E$>$0.1 MeV) of 2.3$\pm$0.7 cm$^{-2}$s$^{-1}$ agrees with the measured 2.16$\pm$0.7 cm$^{-2}$s$^{-1}$ very well. The variation of the $\gamma$-ray flux at different locations on the same level depends on the variation of the rock formations and the radioactivity levels in the rock. This variation can be as large as 30\% as seen in the measurements. This is to say that the $\gamma$-ray flux must be measured in the experimental area where a low-background experiment is to be located.
 
\section{The simulated neutron fluxes for different levels}

Deep underground neutrons come from three primary sources: 1) the  $(\alpha, n)$ neutrons induced by the $\alpha$-decays in $^{238}$U and $^{232}$Th decay chains in the rock, 2) the spontaneous fission decay of  $^{238}$U, and 3) the muon-induced neutrons. The muon-induced neutrons have a strong dependence on the depth~\cite{meihime}. In general the fluxes of the cosmogenic neutrons are one order, two orders, and three orders of magnitude smaller than that of  the radiogenic neutrons at the 800-ft level, the 2000-ft level, and the 4550-ft level, respectively. However, the cosmogenic neutron energy spectrum is much harder~\cite{meihime} and  they are beyond the scope of this work. Neutrons generate $\gamma$ rays through inelastic scattering and capture on nuclei. These $\gamma$ rays have a higher energy than the $\gamma$ rays from radioactive decay described in the above sections. We simulate the neutron propagation in the rock with the same geometry that is used to simulate the $\gamma$ rays described in Fig.~\ref{geometrymap}. The mean free path of the radiogenic neutrons is about 10 cm in the rock. These neutrons can be substantially attenuated in the 3 m rock. Therefore, the contribution of the neutrons beyond 3 m is negligible.  

\subsection{Neutrons from  $(\alpha, n)$ Reactions}

The decays of $^{238}$U, $^{232}$Th, and their daughters produce $\alpha$ particles, $\gamma$ rays, and $\beta$ particles. The $\alpha$ particles with energies on the order of a few MeV range are capable of generating neutrons through $(\alpha, n )$ reactions in the rocks. The thickness of the rock in the simulation is 3 meters.    
\par
The thick target yield of $(\alpha, n )$ neutrons for an $\alpha$-particle projectile is 
calculated using the equations in Refs.~\cite{meichao, heaton} as below:
\begin{equation}\label{yield}
Y_{i} = \int^{R}_{0}n_{i}\sigma_{i}(E)\rm{d}x
      =\frac{N_{A}}{A_{i}}\int^{E_{0}}_{0}\frac{\sigma_{i}(E)}{S^{m}_{i}(E)}\rm{d}E,
\end{equation}
where $R$ refers to a range of the $\alpha$-particle in the materials.
 $n_{i}$ is the number of atoms per unit volume of the $i$-th element, $\sigma_{i}$
is the microscopic $(\alpha, n)$ reaction cross-section for an $\alpha$-particle with
the energy $E$, and $E_{0}$ stands for the initial energy, 
$S^{m}_{i}$ is the mass stopping power of the $i$-th element, $A_{i}$ is the atomic mass of the $i$-th element, and $N_{A}$ is Avogadro's number.
\par
Under the assumption of secular equilibrium, the $^{232}$Th decay chain yields six $\alpha$ particles and the $^{238}$U decay chain produces eight $\alpha$ particles with various energies $E$. The net neutron yield from the decay chains of $^{232}$Th and $^{238}$U can be determined by the sum of the individual yields induced by each $\alpha$, weighted by the branching ratio for each element and weighted by the mass ratio in the rock. The energy attenuation of $\alpha$-particles in the medium is the dominant process for the thick target hypothesis.
Under the assumption that the incident flux of $\alpha$-particles with energy of E$_{j}$ is invariant until the energy is attenuated to zero, for target element $i$, the differential spectra of neutron yield can be
expressed as:
\begin{eqnarray}
Y_{i}(E_{n}) &= &N_{i}\sum_{j}\Phi_{\alpha}(E_{j})
  \int_{0}^{E_{j}}\frac{\rm{d}\sigma(E_{\alpha}, E_{n})}{\rm{d}E_{\alpha}}\rm{d}E_{\alpha} \nonumber \\
   &= &\frac{N_{A}}{A_{i}}\sum_{j} \frac{R_{\alpha}(E_{j})}{S_{i}^{m}(E_{j})}
  \int_{0}^{E_{j}}\frac{\rm{d}\sigma(E_{\alpha}, E_{n})}{\rm{d}E_{\alpha}}\rm{d}E_{\alpha}, \label{diff}
\end{eqnarray}
where $N_{i}$ is the total atomic number of the rock and
$\Phi_{\alpha}(E_{j})$ is the flux of an $\alpha$-particle with the specific energy $E_{j}$.
 $R_{\alpha}(E_{j})$ refers to the $\alpha$-particle production rate
 for the channel with the energy $E_{j}$
from an actinide element decay. If we consider the specific activity of an actinide
with the concentration per ppm per gram per year, then
$$R_{\alpha}(E_{j}) = 10^{-6}\frac{N_{A}}{A_{a}}\frac{\ln2}{t_{1/2}}B_{j}.$$
Where $A_{a}$ stands for the mass number of the actinide element, specifically $^{238}$U or $^{232}$Th here,
 $B_{j}$ represents the $\alpha$-particle
branching ratio for the certain energy decay channel $E_{j}$, and $ t_{1/2}$ is the half life of the decay.
\par
The thick target yield for a compound is discussed in Refs.~\cite{meichao, heaton}. In order to simplify 
the calculation, the compound is assumed to be
a homogeneous mixture and Bragg's law of additivity for stopping power cross sections holds for the compound.
Based on these assumptions and Eq.(\ref{diff}), the yield for a compound $Y_{c}$ can be written as 
\begin{equation}\label{compound}
 Y_{c}= \sum_{i} w_{i}Y_{i},
\end{equation}
where $w_{i}$ is the mass weight of element $i$ in the compound.
\par
The cross section in Eq.(\ref{diff}), e.g.,
 $\int_{0}^{E_{j}}\frac{\rm{d}\sigma(E_{\alpha}, E_{n})}{\rm{d}E_{\alpha}}\rm{d}E_{\alpha} $, is obtained from 
 TALYS nuclear reaction code~\cite{talys} in which the cross sections of neutron production for all possible reaction channels are calculated. 
The flux of $\alpha$-particles is obtained by
combining the production rate of $\alpha$ particles with the corresponding mass stopping power in the target. The mass stopping
power for specific energies is calculated by our simulation
described in Ref.~\cite{meibao} and the ASTAR program~\cite{astar}. We use the decay chains of selected isotopes in Ref.~\cite{dsd} for $\alpha$-particle emission from $^{238}$U and $^{232}$Th decays. Only decays with visible energies larger than 0.1 MeV or branching ratio more than 0.5\%
are included. 

The overall uncertainty in the calculation of the net neutron yield is justified to be less than 20\%. The sources of the uncertainty are the ($\alpha$,n) reaction cross sections and the calculation of $\alpha$ stopping power in rock. The justification is done by comparing the net neutron yield from this work to similar calculations done for SNO~\cite{heaton} and Gran Sasso~\cite{gran}.

\subsection{Neutrons from Spontaneous Fission}

Naturally, neutrons can also be produced by spontaneous fission of the radioactive elements, e.g., $^{238}$U,$^{235}$U, and $^{232}$Th. The spontaneous fission of $^{238}$U  produces the most significant number of fission neutrons due to its shorter fission half-lives. The other fission neutron rates are two orders of magnitude smaller. Therefore, we consider the fission neutrons  primarily from $^{238}$U. The spectrum of the fission neutrons follows the Watt spectrum.
\begin{equation}\label{fission}
N(E) = C\exp (-E/a)\sinh (bE)^{1/2}, 
\end{equation}
where we adopt the Watt spectrum parameter $a = 0.7124$ MeV and $b$ = 5.6405 MeV$^{-1}$
 in the calculation. The production rate of spontaneous fission of $^{238}$U is 
0.52$\pm$0.02  neutrons ppm$^{-1}$g$^{-1}$year$^{-1}$ in the rock~\cite{gran}. The uncertainty is from the measured $^{238}$U decay constant for spontaneous fission~\cite{guedes}. 

\subsection{The Simulated Neutron Flux and Neutron-Induced $\gamma$-Ray Flux}
The neutron flux and neutron-induced $\gamma$-ray flux in the experimental hall is predicted by the GEANT4 simulation package.  The $(\alpha, n)$ induced neutrons are isotropically generated in the rock with the
energy spectrum given above for two different sources of actinide elements.
Eq.(\ref{fission}) is used to generate neutrons produced by $^{238}$U fission decay. 

Table~\ref{1ppm-an-flux} and Table~\ref{1ppm-fission-flux} show the simulated differential energy spectra of neutrons induced by $^{238}$U and $^{232}$Th decay chains. These two Tables provide conventional values to be used in the simulation for predicting neutron-induced background for individual experiments if one knows the radioactivity levels in the rock that surrounds the detector. 
\par
\begin{sidewaystable}
\caption{Neutron flux in the simulated experimental hall via $(\alpha, n)$ reaction due to natural radioactivity using the Homestake rock chemical composition. The rows indicate the energy of neutron in MeV size bins. The columns represent the further subdivision of the data 
in the MeV bins into 0.1 MeV bins. The uncertainty in the calculation is about 20\% that is justified in the text.}
\centering
\begin{tabular}{ l l|ccccccccccc}\hline
\multicolumn{2}{l}{Range (MeV)} &$0- .1$ &$.1- .2$
                                &$.2- .3$ &$.3- .4$ &$.4- .5$ &$.5- .6$
                                &$.6- .7$ &$.7- .8$ &$.8- .9$ &$.9- 1$ &Sum \\
\hline
 $E_{n}$ & Source & \multicolumn{10}{c}{
                                     Neutron Yield (ppm$^{-1}$cm$^{-2}$s$^{-1})$}\\
\hline
&$^{238}$U
&1.8e-6&7.5e-8&5.3e-8&3.9e-8&3.0e-8&3.3e-8&3.0e-8&3.4e-8&2.4e-8&1.7e-8&2.2e-6\\
\raisebox{1.5ex}{$0$}&$^{232}$Th
&6.7e-7&2.6e-8&2.0e-8&1.4e-8&9.9e-9&1.2e-8&1.3e-8&1.3e-8&1.0e-8&6.4e-9&7.9e-7\\
\hline
&$^{238}$U
&1.3e-8&1.9e-8&1.8e-8&1.7e-8&1.7e-8&1.9e-8&1.6e-8&1.2e-8&1.1e-8&8.7e-9&1.5e-7\\
\raisebox{1.5ex}{$1$}&$^{232}$Th
&5.0e-9&7.9e-9&7.2e-9&7.4e-9&5.6e-9&6.5e-9&5.2e-9&5.0e-9&7.6e-9&3.7e-9&6.1e-8\\
\hline
&$^{238}$U
&1.3e-8&7.7e-9&1.1e-8&1.2e-8&1.4e-8&9.8e-9&9.6e-9&8.1e-9&6.2e-9&6.1e-9&9.7e-8\\
\raisebox{1.5ex}{$2$}&$^{232}$Th
&4.9e-9&4.1e-9&6.9e-9&4.3e-9&4.9e-9&4.9e-9&4.0e-9&3.9e-9&3.2e-9&2.4e-9&4.4e-8\\
\hline
&$^{238}$U
&6.8e-9&7.5e-9&6.7e-9&3.0e-9&3.1e-9&1.8e-9&1.6e-9&3.3e-9&2.1e-9&2.7e-9&3.8e-8\\
\raisebox{1.5ex}{$3$}&$^{232}$Th
&2.7e-9&2.9e-9&2.0e-9&1.2e-9&1.8e-9&1.6e-9&1.1e-9&1.1e-9&1.3e-9&1.7e-9&1.7e-8\\
\hline
&$^{238}$U
&2.1e-9&1.8e-9&1.6e-9&1.2e-9&1.3e-9&1.8e-9&8.9e-10&1.6e-9&5.9e-10&1.2e-9&1.4e-8\\
\raisebox{1.5ex}{$4$}&$^{232}$Th
&1.0e-9&1.6e-9&1.1e-9&1.2e-9&1.9e-9&9.6e-10&6.4e-10&5.9e-10&5.9e-10&4.3e-10&1.0e-8\\
\hline
&$^{238}$U
&1.0e-9&1.0e-9&8.9e-10&5.9e-10&3.0e-10&4.4e-10&1.5e-10&8.9e-10&5.9e-10&0.0e+00&5.9e-9\\
\raisebox{1.5ex}{$5$}&$^{232}$Th
&7.0e-10&5.9e-10&3.8e-10&4.8e-10&2.1e-10&3.8e-10&6.4e-10&4.3e-10&1.6e-10&2.1e-10&4.2e-9\\
\hline
&$^{238}$U
&8.9e-10&5.9e-10&3.0e-10&3.0e-10&1.5e-10&4.4e-10&1.5e-10&0.0e+00&1.5e-10&1.5e-10&3.1e-9\\
\raisebox{1.5ex}{$6$}&$^{232}$Th
&1.1e-10&5.4e-10&3.8e-10&2.1e-10&2.1e-10&1.6e-10&1.1e-10&0.0e+00&0.0e+00&2.1e-10&1.9e-9\\
\hline
&$^{238}$U
&1.5e-10&3.0e-10&1.5e-10&0.0e+00&4.4e-10&0.0e+00&4.4e-10&0.0e+00&0.0e+00&3.0e-10&1.8e-9\\
\raisebox{1.5ex}{$7$}&$^{232}$Th
&1.1e-10&0.0e+00&0.0e+00&1.1e-10&0.0e+00&0.0e+00&1.1e-10&1.1e-10&5.4e-11&5.4e-11&5.4e-10\\
\hline
&$^{238}$U
&0.0e+00&0.0e+00&0.0e+00&0.0e+00&0.0e+00&0.0e+00&0.0e+00&0.0e+00&0.0e+00&0.0e+00&0.0e+00\\
\raisebox{1.5ex}{$>8$}&$^{232}$Th
&5.4e-11&1.1e-10&5.4e-11&0.0e+00&0.0e+00&1.6e-10&5.4e-11&0.0e+00&5.4e-11&5.4e-11&5.4e-10\\
\hline
\end{tabular}
\label{1ppm-an-flux}
\end{sidewaystable}

\begin{sidewaystable}
\caption{Neutron flux (ppm$^{-1}$cm$^{-2}$s$^{-1})$ in the simulated experimental hall via $^{238}$U
 fission decay due to natural radioactivity using the Homestake chemical composition. The rows indicate the energy of neutron 
in MeV size bins. The columns represent the further subdivision of the data 
in the MeV bins into 0.1 MeV bins. The uncertainty of about 3\% that is stated in the text.}
\centering
\begin{tabular}{c|ccccccccccc}\hline
$E_{n}$ (MeV) &$0- .1$ &$.1- .2$
                                &$.2- .3$ &$.3- .4$ &$.4- .5$ &$.5- .6$
                                &$.6- .7$ &$.7- .8$ &$.8- .9$ &$.9- 1$ &Sum \\
\hline
0
&9.0e-7&3.4e-8&2.4e-8&1.7e-8&1.4e-8&1.8e-8&1.7e-8&1.7e-8&1.3e-8&8.0e-9&1.1e-6\\
\hline
1
&8.2e-9&1.0e-8&7.7e-9&6.4e-9&1.0e-8&6.8e-9&6.7e-9&7.9e-9&5.5e-9&4.7e-9&7.4e-8\\
\hline
2
&5.1e-9&4.7e-9&7.7e-9&4.7e-9&4.5e-9&3.7e-9&3.8e-9&3.5e-9&2.9e-9&2.4e-9&4.3e-8\\
\hline
3
&1.9e-9&2.9e-9&2.2e-9&1.3e-9&1.3e-9&7.0e-10&9.7e-10&1.3e-9&8.4e-10&1.3e-9&1.5e-8\\
\hline
4
&1.2e-9&9.0e-10&1.1e-9&3.5e-10&9.7e-10&7.7e-10&3.5e-10&6.3e-10&6.3e-10&4.2e-10&7.3e-9\\
\hline
5
&3.5e-10&3.5e-10&3.5e-10&2.8e-10&2.8e-10&5.6e-10&1.4e-10&4.9e-10&0.0e+00&0.0e+00&2.8e-9\\
\hline
6
&7.0e-11&0.0e+00&7.0e-11&0.0e+00&0.0e+00&0.0e+00&2.1e-10&0.0e+00&7.0e-11&1.4e-10&5.6e-10\\
\hline
7
&7.0e-11&1.4e-10&0.0e+00&0.0e+00&0.0e+00&0.0e+00&0.0e+00&0.0e+00&0.0e+00&0.0e+00&2.1e-10\\
\hline
$>8$
&0.0e+00&7.0e-11&7.0e-11&7.0e-11&0.0e+00&0.0e+00&0.0e+00&0.0e+00&0.0e+00&0.0e+00&2.1e-10\\
\hline
\end{tabular}
\label{1ppm-fission-flux}
\end{sidewaystable}

Tables~\ref{sum_neutron} and ~\ref{sum_gamma_1} summarize the integrated neutron flux from $(\alpha, n)$ reactions and fission decays and the $\gamma$-ray flux induced by neutrons. The neutron flux is reported in terms of 
thermal neutrons, slow neutrons and fast neutrons. Note that 
the neutron fluxes reported in Table~\ref{sum_neutron} are for the neutrons with kinetic energy greater than 0.1 MeV before
they transport and scatter into the experimental hall. This means the fluxes of the thermal and slow neutrons and neutron-induced $\gamma$ rays are underestimated.
\begin{table}
\caption{The neutron flux induced by $^{238}$U and $^{232}$Th radioactivity in the simulated experimental hall. The quoted errors are justified in the text.}
\centering
\begin{tabular}{cccc}\hline
&Thermal Neutron &Slow Neutron &Fast Neutron\\
Source &$E_{n}<$1 eV & $E_{n}$ in $[$1 eV, 0.1 MeV$]$&$E_{n}>$0.1 MeV\\
&(ppm$^{-1}$cm$^{-2}$s$^{-1})$ &(ppm$^{-1}$cm$^{-2}$s$^{-1})$ & (ppm$^{-1}$cm$^{-2}$s$^{-1})$\\
$^{238}$U ($\alpha, n)$ &(1.17$\pm$0.23)$\times$10$^{-6}$ & (6.79$\pm$1.36)$\times$10$^{-7}$ &(6.46$\pm$1.29)$\times$10$^{-7}$\\
$^{232}$Th ($\alpha, n)$ &(4.19$\pm$0.84)$\times$10$^{-7}$ &(2.48$\pm$0.50)$\times$10$^{-7}$ &(2.64$\pm$0.53)$\times$10$^{-7}$\\
$^{238}$U fission &(5.73$\pm$0.16)$\times$10$^{-7}$ &(3.26$\pm$0.09)$\times$10$^{-7}$ &(3.04$\pm$0.09)$\times$10$^{-7}$\\
\hline
\end{tabular}
\label{sum_neutron}
\end{table}

\begin{table}
\caption{The $\gamma$-ray flux induced by neutrons in the simulated experimental hall. The quoted errors are justified in the text.}
\centering
\begin{tabular}{ccc}\hline
Source& $\gamma$-ray flux (ppm$^{-1}$cm$^{-2}$s$^{-1})$\\     
$^{238}$U & $(\alpha, n)$& (8.47$\pm$1.69)$\times$10$^{-7}$\\
&$fission$& (3.92$\pm$0.11)$\times$10$^{-7}$ \\
{$^{232}$Th}& $(\alpha, n)$& (3.03$\pm$0.61)$\times$10$^{-7}$\\
\hline
\end{tabular}
\label{sum_gamma_1}
\end{table}

\par
Using the results from Tables~\ref{sum_neutron} and ~\ref{sum_gamma_1}, the neutron and neutron induced $\gamma$-ray fluxes can be estimated by multiplying the
flux with the radioactivity of actinide elements in the rock. There are core samples counted by Ref.~\cite{AlSmith} for different levels at the Homestake Mine. Utilizing the highest concentration of radioactivity counted for the 4850-ft level, i.e.,  $^{238}$U: 0.55 ppm, $^{232}$Th: 0.3 ppm and $^{40}$K: 2.21\%; the prediction for the 4850-ft level is shown in Table~\ref{4850ft}. For the 7400-ft level, the highest measured concentrations~\cite{AlSmith} are 
$^{238}$U: 0.49 ppm, $^{232}$Th: 0.20 ppm and K: 0.57\%; giving a neutron yield and $\gamma$-ray flux in Table~\ref{4850ft}.
\begin{table}
\caption{Predicted $\gamma$-ray and neutron fluxes for the 4850-ft and 7400-ft levels. The quoted errors are justified in the text.}
\centering
\begin{tabular}{cccc}\hline
Depth&\multicolumn{2}{c}{$\gamma$-ray flux (cm$^{-2}$s$^{-1}$)}&Neutron flux ( cm$^{-2}$s$^{-1}$)\\ 
     &$^{238}$U and $^{232}$Th&$^{40}$K& \\    
4850 ft& 0.32$\pm$0.10&1.46$\pm$0.44&(2.3$\pm$0.8)$\times$10$^{-6}$\\
7400 ft& 0.27$\pm$0.08&0.38$\pm$0.11&(2.0$\pm$0.7)$\times$10$^{-6}$\\
\hline
\end{tabular}
\label{4850ft}
\end{table}

\section{Discussion and conclusion}
Using three identical NaI detectors, we have investigated the $\gamma$-ray fluxes at the Homestake Mine at several levels that are currently accessible. A simple algorithm is developed to convert the measured $\gamma$-ray rates into a flux.  The measured $\gamma$-ray fluxes are compared to a Monte Carlo simulation by applying known knowledge about the rock chemical composition, rock formation, and radioactivity levels. A good agreement is achieved between the measurements and Monte Carlo simulation. The Monte Carlo simulation is then applied to predict the neutron fluxes and $\gamma$-ray fluxes induced by natural radioactivity for different levels. These results are important for the initial set of physics experiments to be performed at the Sanford Lab. The predictions are useful to the planned low-background experiments at DUSEL. 

\section{Acknowledgment}
The authors wish to thank Bill Roggenthen, Yuen-Dat Chan, Steve Elliott, Vince Guiseppe, and Christina Keller for a careful reading of this manuscript. We thank Jaret Heise and the staff at the Sanford Laboratory for the invaluable support that made this work successful.  Keenan Thomas was also partially supported by the South Dakota Space Grant Consortium. This work was supported by the NSF Grant PHY-0758120.


\begin{thebibliography}{99}
\bibitem{sep} Steve R. Elliott and Petr Vogel, Annu. Rev. Nucl. Part. Sci. 52 (2002) 115.
\bibitem{sej} Steven R. Elliott and Jonathan Engel, J. Phys. G: Nucl. Part. Phys. 30, (2004) R183.
\bibitem{fky} F.T Avignone III, G.S. King III and Yuri Zdesenko, New J. of Phy. 7, (2005) 6.
\bibitem{asb} A.S. Barabash, Physics of Atomic Nuclei 67 No. 3  (2004) 438.
\bibitem{gju} G. Jungman et al., Physics Reports  267 (1996) 195.
\bibitem{mwg} M. W. Goodman and E. Witten, Phys. Rev. D  31 (1985) 3059.
\bibitem{bwl} B. W. Lee and S. Weinberg, Phys. Rev. Lett. 39 (1977) 165.
\bibitem{meichao}D.-M. Mei, C. Zhang, A. Hime, NIM A  606 (2009) 651-660.
\bibitem{heaton} R.K. Heaton, H.W. Lee, B.C. Robertson, E.B. Norman, K.T. Lesko, B. Sur, NIM A 364 (1995) 317.
\bibitem{sno} Perillo Isaac et al., Measurements of the Gamma-Ray Flux at the Sudbury Neutrino Observatory, American Physical Society, Division of Nuclear Physics Meeting, October 2-5, 1996, abstract \#ED.07.
\bibitem{gran} H. Wulandari, J. Jochum, W. Rau, and F. Von Feilitzsch, Astroparticle Physics, V22, (2004) 313-322.
\bibitem{guedes} S. Guedes et al., Journal of Radioanalytical and Nuclear Chemistry 245 (2000) 441-442.
\bibitem{bell} P. Belli et al., Il Nuovo Cimento A, V 101, (1989) 959-966.
\bibitem{frej} H. Ohsumi et al., Nucl. Instr. and Meth. in Phys. A 482 (2002) 832-839.
\bibitem{boulby} Vitaly A. Kudryavtesev, 10th Int. Conf. on Topics in Astroparticle and Underground Physics (TAUP2007),  doi:10.1088/1742-6596/120/4/042028.
\bibitem{baksan} J. N. Abdurashitov et al., Physics of Atomic Nuclei, V63, (2000) 1276-1281.
\bibitem{xia} X-Ray Instruments Associates, 8450 Central Avenue, Newark CA 94560, USA.
\bibitem{igor} Wavemetrics Inc., PO Box 2088, Lake Oswego, OR 97035, USA.
\bibitem{root} An object oriented framework for large scale data analysis developed at CERN. http://root.cern.ch/drupal/.
\bibitem{AlSmith} Al Smith, LBL, Homestake (SUSEL/DUSEL) Samples Results of Radiometic Analysis at LBNL, Private communication.
\bibitem{geology} W. Roggenthen and A. R. Smith, Uranium, Thorium, Potassium Contents of Materials in the Homestake Underground, Lead SD, 2008 AGU Fall Meeting, San Francisco, CA.
\bibitem{attu} http://physics.nist.gov/PhyRefData/Xcom/Text/XCOM.html.
\bibitem{meikeenan} K. Thomas et al., Radon Measurements at the Sanford Laboratory at Homestake (white paper).
\bibitem{rocksample} Jordan, B. T., 2009, Geochemistry tectonic setting of the Yates unit of the Poorman Formation (DUSEL bedrock) and other northern Black Hills amphibolites: Geological Society of America Abstracts with Programs, v. 41, n. 7, p. 271.
\bibitem{types} H. Rogers, Geology of Precambrian rocks in the  Poorman anticlinorium and Homestake mine, Black Hills, South Dakota, in Metallogeny of gold in the Black Hills, South Dakota, eds. C. J. Paterson, and A. L. Lisenbee, Guidebook Prepared for Soc. Econ. Geol. Field Conf. 5-9 September, 1990.
\bibitem{tom} Private conversation with T. C. Trancynger, P. G. Science Liaison Supervisor with a M.S. degree in geology and professional geologist certification, who has been working at Homestake Mine for more than 30 years.
\bibitem{meihime} D.-M. Mei and A. Hime, Phys. Rev. D 73 (2006) 053004.
\bibitem{branch} http://www.euronuclear.org/info/encyclopedia/d/decaybasinnatural.htm.
\bibitem{geant4} S. Agostinelli et al., Nucl. Instr. and Meth. in Phys. A 506 (2003) 250-303. J. Allison et al., IEEE Transactions on Nuclear Science 53 (2006) 270-278.
\bibitem{talys} A.J. Koning, S. Hilaire, M.C. Duijvestijn, TALYS: comprehensive nuclear reaction modeling, in: R.C. Haight, M.B. Chadwick, T. Kawano, P. Talou (Eds.), Proceedings of the International Conference on Nuclear Data for Science and Technology-ND2004, AIP, vol. 769, September 26–October 1, 2004, Sante Fe, USA, 2005, p. 1154.
\bibitem{meibao} D.-M. Mei, et al., Astropart. Phys. 30 (2008) 12.
\bibitem{astar} http://physics.nist.gov/PhysRefData/Star/Text/ASTAR.html.
\bibitem{dsd} D.S. Delion, A. Insolia, R.J. Liotta, Nucl. Phys. (Suppl.) A 654 (1999) 673c; A.M. Sanchez, P.R. Montero, Nucl. Instr. and Meth. A 420 (1999) 481.

\end{thebibliography}
\end{document}